\documentclass[11pt,a4paper]{article}
\pdfoutput=1
\usepackage{jheppub}
\usepackage{amsmath,amscd}
\usepackage{amsfonts}
\usepackage{amssymb}
\usepackage{graphicx}
\usepackage{color}
\usepackage[normalem]{ulem}
\usepackage{hyperref}
\usepackage{enumerate}
\usepackage{mathtools}
\usepackage{slashed}
\usepackage[symbol]{footmisc}

\newcommand{\CC}{\mathbb{C}} 
\newcommand{\RR}{\mathbb{R}} 
\newcommand{\ZZ}{\mathbb{Z}} 
\newcommand{\NN}{\mathbb{N}} 

\newcommand{\G}{\mathcal{G}}

\newcommand{\T}{\mathcal{T}}

\def\tr         {{\rm  tr}}
\def\cala         {{\cal A}}

\def\calf         {{\cal F}}
\def\calg         {{\cal G}}

\def\calh         {{\cal H}}
\def\cali         {{\cal I}}

\def\call         {{\cal L}}
\def\calm         {{\cal M}}
\def\caln         {{\cal N}}
\def\calo         {{\cal O}}

\def\calq         {{\cal Q}}

\def\cals         {{\cal S}}

\def\be{\begin{equation}}
\def\ee{\end{equation}}
\def\bea{\begin{eqnarray}}
\def\eea{\end{eqnarray}}

\def\a{\alpha}
\def\b{\beta}

\def\g{\gamma}
\def\G{\Gamma}
\def\d{\delta}
\def\e{\epsilon}
\def\D{\Delta}
\def\l{\lambda}

\def\m{\mu}
\def\x{\xi}

\def\o{\omega}
\def\O{\Omega}
\def\p{\pi}
\def\P{\Pi}
\def\r{\rho}
\def\z{\zeta}
\def\s{\sigma}

\def\t{\tau}

\def\sF{{{ F}\!\!\!\!\hskip.8pt\hbox{\raise1pt\hbox{/}}\,}}
\def\som{{{ \omega}\!\!\!\!\hskip.8pt\hbox{\raise1pt\hbox{/}}\,}}
\def\sJ{{{\rm J}\!\!\!\!\hskip.8pt\hbox{\raise1pt\hbox{/}}\,}}

\def\pa{\partial}
\def\T{\tau}

\def\to{\rightarrow}
\def\nonu{\nonumber \\{}}
\def\half{{1 \over 2}}

\title{Superconformal indices and localization in $N=2B$ quantum mechanics}
\author[a]{Joris Raeymaekers,}
\author[a]{Canberk \c{S}anl{\i}}
\author[b,c]{and Dieter Van den Bleeken\footnote[1]{Currently at Department of Meteorological and Climate Research, Royal Meteorological Institute, 1180 Uccle, Belgium.}}

\affiliation[a]{CEICO, Institute of Physics of the Czech Academy of Sciences,\\  Na Slovance 2, 182 00 Prague 8, Czech Republic.}
\affiliation[b]{Physics Department, Boğaziçi University\\
	34342 Bebek / Istanbul, Turkey}
\affiliation[c]{Secondary address:\\
	Institute for Theoretical Physics, KU Leuven\\
	3001 Leuven, Belgium}

\emailAdd{joris@fzu.cz}
\emailAdd{sanli@fzu.cz}
\emailAdd{dieter.vdbl@gmail.com}

\abstract{Superconformal `type B' quantum mechanical sigma models arise in a variety of interesting contexts, such as the description of  D-brane bound states in an AdS$_2$ decoupling limit. 
Focusing on $N=2B$ models,	we  study  superconformal indices which count short  multiplets  and provide an alternative to the standard Witten index, as the latter suffers from infrared issues. 
	We show that the basic index receives contributions from lowest Landau level states in an effective magnetic field and that,  due to the noncompactness of the target space, it is typically divergent. 
	Fortunately, the  models of interest  possess an additional target space isometry which allows for the definition of a well-behaved refined index. We compute this  index using  localization of the functional integral and find that the result agrees with    a naive application of the Atiyah-Bott fixed point formula outside of it's starting    assumptions.
In the simplest examples, this formula can also be directly verified by explicitly computing the short multiplet spectrum.}
	
\arxivnumber{}
\keywords{}
\begin{document}
 \maketitle

\section{Introduction}
\renewcommand*{\thefootnote}{\arabic{footnote}}
Conformally invariant quantum mechanical theories, and their supersymmetric extensions, govern a variety of physically interesting systems, from instanton moduli spaces to the microscopic structure of
extremal black holes. The study of such models goes back to  \cite{deAlfaro:1976vlx},  and we refer to \cite{BrittoPacumio:1999ax,Fedoruk:2011aa} for reviews and a guide to the literature.  For superconformal systems with extended supersymmetry, it is possible to define a superconformal index \cite{Fubini:1984hf} which is a generalized Witten index \cite{Witten:1982df} of the form
\be 
\O_\calf [h] = \tr (-1)^\calf e^{ - \b \{\calg, \calg^\dagger \}  } \hat h.\label{indintro}
\ee
Here, $\calg$ is a combination of Poincar\'e and conformal supercharges,  and the operator $\calf$ can be a fermion number operator or a suitable R-charge. We allow also for a possible refinement by an additional element  $\hat h$ which commutes with these operators.  
The index $\O_\calf [h]$  captures crucial 
information on the short multiplet spectrum of the theory and has the usual properties of an index, in that  it is independent of $\b$  and should be computable using localization methods.  Furthermore, (\ref{indintro}) provides a natural definition of  $\tr (-1)^\calf \hat h$ in such theories, since if $\calg$ were taken to be  one of the Poincar\'e supercharges the index would suffer from subtle infrared issues due to the gapless continuous spectrum of the Hamiltonian  in scale-invariant theories.



In an inspiring recent series of papers \cite{Dorey:2018klg,DoreyC,Dorey:2019kaf}, Dorey and collaborators computed   superconformal indices in a class of quantum mechanical sigma models. They focused on $N$-extended `type A' sigma models,  which are sometimes referred to as $(N/2,N/2)$ supersymmetric as they can be obtained by dimensional reduction from 1+1  dimensional sigma models of this type.   For $N=4$ these models are characterized by a target space\footnote{For an overview of   supersymmetric quantum mechanical sigma models, see \cite{Smilga:2020nte}.}  which is  K\"ahler, and for higher $N$  it possesses additional geometric structure, for example hyperK\"ahler when $N=8$. Since in most physical applications the target space is in fact  singular, the authors proposed a definition of the index  using  a resolution of the target space, in the cases when it can be described as an symplectic complex variety, and used  localization theorems to compute it.


Localization methods have so far not been applied to the computation of the superconformal index in $N$-extended `type B' sigma models\footnote{See however Appendix B of \cite{Benini:2015eyy} for a direct computation of the index in a simple example.}. These are sometimes referred to as $(0,N)$ models   after their 2-dimensional parent theories. 
The target spaces possess somewhat less familiar geometric structures  such as `K\"ahler with torsion' for $N=2$ and `hyperK\"ahler with torsion' for $N=4$.

 Type B sigma  models are nevertheless of significant interest as  they   arise in the description of D-brane systems and black holes in string theory  \cite{Michelson:1999zf,Michelson:1999dx,Britto-Pacumio:2000fnm}. In this context, they are expected to play a role in a top-down understanding  of AdS$_2$/CFT$_1$ duality \cite{Sen:2008yk,Bena:2018bbd}. One example which motivates the present study is that  of the Coulomb branch of the quiver  mechanics governing bound states of D-branes  \cite{Denef:2002ru}, which can be described as an $N=4$ type B sigma model \cite{Mirfendereski:2020rrk,Mirfendereski:2022omg}.  It  develops an enhanced $D(2,1; 0)$ superconformal  symmetry in the deep scaling regime in which an AdS$_2$ black hole throat forms \cite{Anninos:2013nra,Mirfendereski:2020rrk}. Here, it is hoped that the superconformal index could  serve as a regularized version of the standard BPS index  (see e.g. \cite{deBoer:2008zn,deBoer:2009un,Manschot:2011xc,Hori:2014tda}) and shed light on the fate of `pure Higgs' states \cite{Bena:2012hf,Manschot:2012rx}   in this subtle regime. 
 Another motivation comes from the   description of asymptotically AdS$_4$ black holes in M-theory which also possess an AdS$_2$ near-horizon region.  Microscopic aspects  are expected \cite{Benini:2015eyy} to be captured by an $N=2B$   superconformal sigma model, albeit in the presence of additional multiplets of  a different (Fermi) type (see also \cite{Bullimore:2019qnt,Benini:2022bwa,Benini:2024cpf}). The geometry of $\caln=2B$ superconformal mechanics with additional Fermi multiplets is known \cite{Papadopoulos:2000ka}, and can be used \cite{chenwip} to compute (\ref{indintro}) as a first step towards understanding $AdS_4$ black hole entropy, and thereby obtaining a relation to giant graviton expansions \cite{Choi:2022ovw,Arai:2020uwd}, from a CFT$_1$ perspective.

In this work we initiate the study of superconformal indices 
in type B sigma models with at least $N=2$ supersymmetry, which as we shall explain is the minimum for which a sensible superconformal index can be defined. 
The models of interest possess an $su(1,1|1)$ superconformal symmetry and the index counts the short multiplets weighted by  their $(-1)^\calf \hat h$ parity. The computation of the superconformal index presents two difficulties compared  to that of the standard Witten index for compact sigma models: firstly, the target space is necessarily {\em non-compact} and, secondly, that the geometry in many  physically interesting models  takes the form of a {\em singular cone}. In this work, we will focus on the first issue, studying  first explicitly the subclass of models with regular target spaces (which are necessarily flat). For singular target spaces we will proceed under the assumption that the target space can be resolved in a way that preserves the  $N=2$ subalgebra appearing in the definition of the index (\ref{indintro}). 
That is, we will  consider the computation  of   (\ref{indintro}) in more general $N=2B$ supersymmetric sigma models with regular but non-compact target spaces, leaving the details of the  resolution  mechanism\footnote{In the subclass of models where the target space is  Ricci-flat and K\"ahler,  resolutions of the singularity were studied in detail in \cite{Martelli:2005tp}.} for future study \cite{p2}.

We find that, similar to what happens for the type 4A models  \cite{Dorey:2018klg},
  the sigma model that appears in the computation of (\ref{indintro}) involves
  an effective background magnetic field, and the index receives contributions  from its lowest Landau levels\footnote{The relation between superconformal chiral primaries and lowest Landau levels also played a crucial role in the older works \cite{Gaiotto:2004ij,Raeymaekers:2005ig,Denef:2007yt}.}. The number of lowest Landau level states in our non-compact target space is typically infinite, so that the unrefined index with $\hat h=1$   is either divergent or an indeterminate alternating infinite sum, depending on how one chooses  the operator $\calf$. Fortunately, we will show that this issue can be remedied, as our models always possess an additional charge $J$  coming from an additional Reeb-like isometry of the target space. 
We will argue that the  refined index (\ref{indintro}) with $\hat h = \z^J$ is  well-behaved   since the number of states with fixed $J$-charge is finite in our models.
The refined index can  formally be  interpreted  as the character-valued index of an appropriate elliptic complex, which on a compact  manifold would be given by   the Atiyah-Bott fixed point formula \cite{Atiyah-Bott}. One of our main points is that this formula can still be applied in the current non-compact situation, and follows from a path integral derivation using supersymmetric localization. For this purpose we need a slight generalization of Alvarez-Gaum\'e's proof \cite{Alvarez-Gaume:1983zxc} (see \cite{Alvarez-Gaume:1986ggp} for a pedagogical introduction) of the character-valued index theorem which includes the effective background magnetic field and relaxes the assumption of K\"ahlerity. For regular target spaces, we can compare our localization results against the direct computation of the index from the construction of the contributing BPS states, as   in Appendix 2 of \cite{Benini:2015eyy}.

  This work is structured as follows. In Section \ref{Secsuconfalg} we discuss some generalities of superconformal quantum mechanical theories. We focus on the minimal setup with $N=2B$ symmetry, for which we define various superconformal indices and their representation-theoretic content. In Section \ref{Secgensigma} we discuss superconformal indices in general superconformal  $N=2B$  sigma models  and propose a tentative  index theorem to compute them. In Section \ref{Secfreepart} we verify this proposal in the simplest examples with regular target spaces, both using explicit computation as well as using a localization argument. In Section \ref{SecPI} we generalize this localization computation to general (resolved) superconformal sigma models.


\section{The $N=2$ superconformal algebra and indices}\label{Secsuconfalg}
In this section we collect some general remarks on the superconformal index in quantum mechanics.  We will see that a sensible superconformal index needs at least $N=2$ extended supersymmetry, and we focus on this minimal case. We  discuss superconformal indices and  the information about the short multiplet spectrum they capture, as well as a useful criterion under which a refined index exists.
\subsection{Superconformal index in quantum mechanics}
Our interest in this work is in conformally invariant quantum mechanical models, whose study goes back to \cite{deAlfaro:1976vlx},  see e.g. \cite{BrittoPacumio:1999ax,Fedoruk:2011aa} for reviews. These models possess, besides a Hamiltonian $H$, {also} a dilatation operator $D$ and special conformal generator $K${, which are selfadjoint operators on the Hilbert space, and whose commutation relations form the $sl(2,\RR)$ algebra}
\begin{equation}
[D,H]=-iH\qquad [D,K]=i K\qquad [ H, K]=2i D.\label{sl2r1}
\end{equation}
We should note that {this implies that} the spectrum of $H$ is continuous since acting with $e^{i \l D}$ rescales energies with a factor $e^{{-}\l}$. 
We will focus on  theories {that include fermions and} in which the symmetry is enhanced to a superconformal algebra. Models with $N$-extended superconformal symmetry possess additional odd  generators: the supersymmetry charges $Q^\a$ and superconformal charges $S^\a, \a = 1, \ldots , N$, satisfying
\be 
	\{Q^\a,Q^\b\}=2\delta^{\a\b}H,\qquad \{S^\a,S^\b\}=2 \delta^{\a\b}K\,.\label{suconfN}
\ee
A classification of the superconformal algebras at various values of $N$ can be found {e.g. in \cite{Claus1998}.}

We are interested in computing generalizations of the  supersymmetric Witten index \cite{Witten:1982df}, which contain information on the spectrum of short superconformal multiplets in the theory. 
For this purpose we need to specify the following ingredients:
\begin{itemize}
	\item A subalgebra of the superconformal algebra (\ref{sl2r1},\ref{suconfN}) based on a fermionic generator $\calg$,
	\be 
	\{ \calg, \calg^\dagger \} =\calh, \qquad [\calg, \calh ] =0.
	\ee
	Here, $\calh$  is the Hamiltonian of an auxiliary supersymmetric quantum mechanical system and will typically be different from $H$ in (\ref{sl2r1},\ref{suconfN})
	\item A grading operator $(-1)^\calf$, where  $\calf$ is a selfadjoint bosonic operator with an integer spectrum under which $\calg$ has odd charge. In practice, $\calf$ will be either a fermion number operator or, similar to the BPS index \cite{Denef:2007vg},  an appropriate $R$-charge.
	\item Optionally, in order to refine the index,  we may specify an additional operator $\hat h$ with even $\calf$-charge and which commutes with $\calg$ and $\calg^\dagger$.
 \end{itemize}
We then consider a  general index of the form
\be 
\O_\calf [h] = \tr (-1)^\calf e^{ - \b \calh } \hat h .\label{indf}
\ee
In order for this object to have the usual nice properties of an index, we will impose the standard \cite{2012supersymmetric} additional requirement that 
the Hamiltonian $\calh$ has a gapped spectrum. This avoids infrared issues and ensures that  $\O_\calf [h]$ 
is independent of  $\b$. 
Naive computations of the index for a non-gapped $\calh$ can lead to inconsistent results \cite{akhoury1984anomalous,2012supersymmetric} and properly  defining the index is in this case a  highly subtle problem    \cite{Lee:2016dbm}   which we will sidestep in this work.

Returning now to superconformal theories (\ref{suconfN}), we see that the naive supersymmetric Witten index, where $\calg$ is taken to be one of the supercharges $Q^a$ and $\calh$ to be the Hamiltonian $H$, fails the criterion of being gapless, since as we remarked above it has a continuous spectrum extending down to zero. 
To remedy the situation, one can try to obtain a well-defined index by taking $\calg$ to be a combination of supercharges and superconformal charges; in that case we call $\O_\calf [h]$  a superconformal index. As emphasized in \cite{Britto-Pacumio:2000fnm}, the superconformal index can be viewed as a physical  definition of the  infrared-subtle Witten index  with $\calh = H$ in such theories. 

\subsection{The algebra $su(1,1|1)$}
 As we {discuss} in Appendix \ref{AppnoindNis1}, superconformal indices satisfying the above criteria can exist only for theories with $N >1 $, and we will restrict attention to the simplest case $N=2$ in this work.
The unique $N=2$ superconformal algebra \cite{Claus1998} is $su(1,1|1)=osp(2|2)=sl(2|1)$, with bosonic subalgebra  $sl(2,\mathbb{R}) \oplus   u(1)$. Denoting the $u(1)$ generator as\footnote{Note that the generator $R$ is differently normalized from the one in \cite{Mirfendereski:2022omg}: $R= -2 R_{\rm there}$.} $R$, the \mbox{(anti-)} commutation relations supplementing (\ref{sl2r1}) and  (\ref{suconfN}) are 
\bea
\,	[D,Q^\a]&=&-\frac{i}{2}Q^\a\qquad [D,S^\a]=\frac{i}{2}S^\a\qquad [H,S^\a]=-i \,Q^\a\qquad [K,Q^\a]=i S^\a\label{slsusy2}\\
\,	[R,Q^\a]&=&i\epsilon^{\a\b}Q^\b\qquad [R,S^\a]=i\epsilon^{\a\b}S^\b\\
\, \{Q^\a,S^\b\}&=&-2\delta^{\a\b}D+\epsilon^{\a\b}R
\eea
Defining the complex combinations
\be
\calq \equiv\half( Q^1 + i Q^2),\qquad  \cals \equiv \half ( S^1 + i S^2),
\ee
it will be useful to make a change of basis, depending on a continuous ans positive parameter 
$\o$ which  (in units where $\hbar = 1$) has the dimension of a frequency: 
\bea 
L_0&=&\frac{1}{2}(\omega^{-1} H+\omega K)\label{vir1}\\
L_{\pm 1}&=&\frac{1}{2}( \omega^{-1}H-\omega K)\pm i D\label{vir2}\\
\calg_{\pm \half} &=& \omega^{-\frac{1}{2}}\calq \mp i\omega^{\frac{1}{2}} \cals .\label{calgdefs}
\eea
The algebra becomes, in the new basis,
\begin{align}\label{algvir}
{} [ L_m, L_n]&=(m-n) L_{m+n}& & & & \\
{} [L_0,\calg_{\pm \half}]&=\mp\frac{1}{2}\calg_{\pm \half}\label{rl}, & [L_{\mp 1},\calg_{\pm \half}]&=\pm \calg_{\mp \half}, & 
[R,\calg_{\pm \half}]&=\calg_{\pm \half}\\
\{\calg_{\pm \half},\calg_{\pm \half}^\dagger\}&=2L_0\pm R\label{BPS}, &
\{\calg_{\pm \half},\calg_{\mp \half}^\dagger\}&=2 L_{\pm 1}, &
\{\calg_{\a },\calg_{\b}\}&= 0.
\end{align}
An important identity for what follows relates $\calg_{ 1/2}$ and  $\calg_{- 1/2}$ to $\calq$ through simple similarity transformations:
\be
\calg_{\pm \half} =    \o^{-\half}e^{\mp \o K} \calq  e^{\pm \o K}.\label{calgcalq}
\ee
The operator $L_0$ is related to the dilatation operator by the similarity transformation 
\be 
L_0= i M_\o^{-1} D M_\o , \qquad  M_\o = e^{ - { H \over 2 \o}} e^{\o K}.
\ee
Therefore, working in the new basis can roughly be thought   of as organizing the Hilbert space in eigenspaces of the dilatation generator whose spectrum is discrete \cite{deAlfaro:1976vlx}.
One can also show that
\be
L_0 =  U_\o^\dagger (L_0)_{|\o=1} U_\o , \qquad  U_\o = e^{ i {\rm arctanh} {\o^2-1 \over \o^2+ 1} D}.\label{L0omunit}
\ee
In other words, 
 the $L_0$ operators at different values of $\o$ are in fact unitarily equivalent, except in the limits $\o \to 0$ and $\o \to \infty$, where the   operator $U_\o$ ceases to be well-defined. We will therefore set $\o =1$ in what follows.

\subsection{Lowest weight representations}
Let us also comment on the  representations of this algebra relevant for our purposes. These are of `lowest  weight' type  with respect to $L_0$. {Starting from} a primary $| h, r \rangle$, where $h$ and $r$ are $L_0$  {respectively} $R$ eigenvalues, which is annihilated by the lowering operators $\calg_{1/2}, \calg_{-1/2}^\dagger$ and $L_1${, the multiplet is built using the raising operators $\calg_{-1/2}, \calg_{1/2}^\dagger$ and $L_{-1}$}. From (\ref{BPS})  {one derives} the unitarity bound
\be 
2 h \geq |r| \label{unbound}
\ee
which, when   saturated,  leads to a short multiplet. When  $ r = 2 h$  the short multiplet is called  chiral. It is built on  a chiral  primary state $|\chi \rangle$ which is annihilated by $\calg_{-1/2}$ and $\calg_{-1/2}^\dagger$. This requirement is sufficient, since such a state is automatically annihilated by the remaining lowering operators $\calg_{1/2}^\dagger$ and $L_1$.  Indeed, if $ \calg_{1/2}^\dagger |\chi \rangle$ were nonzero, then it would have a negative eigenvalue under $\calh_-$, in conflict with this operator's positivity. Invariance under $L_1$ then follows from the algebra (\ref{algvir}).
 Similarly, for  $ r = - 2 h$ we speak of an antichiral multiplet,    which is built on an antichiral  primary state  $|\tilde \chi \rangle$ annihilated by $ \calg_{1/2}$ and $\calg_{1/2}^\dagger$. Summarizing:
     \begin{align}
   &{\rm chiral\ primary:}  & \calg_{-1/2}|\chi \rangle= \calg_{-1/2}^\dagger|\chi \rangle =0 \nonu
   &{\rm anti-chiral\ primary:} & \calg_{1/2}|\tilde \chi \rangle= \calg_{1/2}^\dagger|\tilde \chi \rangle =0 \label{shortcontsum}
   \end{align}
   The corresponding short multiplets have the following representation content under the  bosonic subgroup:
\begin{align}
&{\rm chiral:}  &(h)_r + \left(h+ \half\right)_{r-1}, & &r = 2h,\nonu
&{\rm anti-chiral:} & (h)_r + \left(h+ \half\right)_{r+1}, & &r = -2 h,\label{shortcont}
\end{align}
where $(h)_r$ denotes the $sl(2, \RR )$ representation built on the primary $|h,r\rangle$ by repeated application of $L_{-1}$.
When $ 2 h > |r|$, we get a long multiplet whose bosonic content is  {the combination} of a chiral and an antichiral multiplet:
\begin{align}
	&{\rm long:}  &(h)_r + \left(h+ \half\right)_{r-1}+ \left(h+ \half\right)_{r+1}+ \left(h+ 1\right)_{r}, & &|r| < 2h .\label{longcont}
\end{align}

\subsection{Superconformal indices}
We now { identify the ingredients necessary for defining a superconformal index of the form (\ref{indf}).} The  two natural choices for the fermionic  generator $\calg$ are  $\calg = \calg_{ 1/2}$  and $\calg = \calg_{- 1/2}$ and lead to an index receiving contributions from antichiral  resp. chiral primaries.
The corresponding dimensionless `Hamiltonians' are 
\be 
\calh_\pm  =  \{ \calg_{\pm \half}, \calg_{\pm \half}^\dagger \} =  2 L_0 \pm R .
\ee
{The operator $\calh_\pm$ will always have a discrete spectrum in  models that, as an {$su(1,1|1)$} representation, decompose as a {discrete} sum of lowest weight representations. As we will see below these include a large class of superconformal sigma models.} We therefore expect $\calh_\pm$ to satisfy our requirement of being gapped. Leaving the possibility of refining the index aside for the moment, we will consider the  superconformal indices 
\be 
\O_\calf^\pm = \tr (-1)^\calf e^{ - \b \calh_\pm }.\label{indunref}
\ee
For the grading operator $\calf$, there will also  be two natural choices: we could take it to be a fermion number operator $F$ or the R-charge operator $R$, if the latter has an integer spectrum\footnote{The choice $\calf = R$ is the one considered in  \cite{Britto-Pacumio:2000fnm}, while \cite{Dorey:2018klg} considered $\calf = F$.}. 
 By construction these indices count (anti-) chiral multiplets weighted by the $(-1)^\calf$ parity of their ground states as follows:
\be
\O^+_\calf =\sum_h \left(  N_{\rm anti-chiral}^{\calf\ \rm even} (h) - N_{\rm anti-chiral}^{\calf\ \rm odd} (h)\right) , \qquad \O^-_\calf =\sum_h \left(  N_{\rm chiral}^{\calf\ \rm even} (h) - N_{\rm chiral}^{\calf\ \rm odd} (h)\right).\label{indmult}
\ee
As a consistency check, it is straightforward to evaluate $\O_\calf^+$ on the different types of multiplet, using the decompositions (\ref{shortcont},\ref{longcont}) and expressions for the bosonic characters,  and show that chiral and  long multiplets  do not contribute to $\O_\calf^+$ , while the anti-chiral multiplet contributions lead to  (\ref{indmult}) (and similarly for $\O^-_\calf$).

Let us move to  the possibility of defining refined  superconformal indices  which will play an important role in what follows. A refined index contains more information about the spectrum by keeping track of an additional quantum number  $J$, and depends on the corresponding chemical potential. 
This will be especially relevant for our the sigma models we are about to consider, as we shall see that due to noncompactness of the target space an infinite number of states contribute, leading to unrefined indices which are infinite or ill-defined alternating sums of 1's and -1's. The refined index on the other hand will be well-behaved as long as the number of states at fixed $J$-charge   is finite.     To obtain such a refinement the selfadjoint  operator $J$ should commute  with  $ (-1)^\calf$ and  with $\calg_{1/2}$ (or 
$\calg_{-1/2}$) and not be proportional to $\calh_+$ (or $\calh_-$). To define the refined index, we set $\hat h =\z^{\pm J}$ in (\ref{indf})  
\be 
\O^{\pm}_\calf [\z ] = \tr (-1)^\calf e^{- \b \calh_\pm} \z^{\pm J}.\label{indref}
\ee
The signs in the exponent of $\z$ are chosen for later convenience (to ensure, in our conventions,  convergence for $|\z|<1$). 

 It is clear from (\ref{BPS}) that no suitable operator $J$ exists  within the $su(1,1|1)$ algebra itself. However,  most models allow in addition for the definition of a fermion number operator $F$ which is independent of the R-charge $R$.  
If $F$ can be chosen to assign fermion number one to both $\calg_{1/2}$ and $\calg_{-1/2}$, namely
\be 
[ F, \calg_{\pm \half}] = \calg_{\pm \half},\label{GsFis1}
\ee
then it is straightforward to see that the difference $F-R$ can be used to refine the index. If (\ref{GsFis1}) holds,  the operator
\be 
J = F - R + c, \label{defJ}
\ee
with $c$ a   c-number setting the zero point, 
commutes with all $su(1,1|1)$ generators
and extends the symmetry to $su(1,1|1) \oplus u(1)$. We can then use $J$ to define the refined indices (\ref{indref}), which  
 have the following interpretation in terms of the short multiplet spectrum:
 \bea 
\O^+_\calf [ \z] &=&\sum_{h,j} \left(  N_{\rm anti-chiral}^{\calf\ \rm even} (h, j) - N_{\rm anti-chiral}^{\calf\ \rm odd} (h,j)\right)\z^j,\nonu
\O^-_\calf [ \z] &=&\sum_{h,j} \left(  N_{\rm chiral}^{\calf\ \rm even} (h, j) - N_{\rm chiral}^{\calf\ \rm odd} (h,j)\right)\z^{-j}.\label{indrefmults}
\eea

Let us also remark that the refined indices defined with $\calf = F$ and with $\calf = R$ are not independent; indeed from (\ref{defJ}) we have
\be 
\O^\pm_F [- \z] = e^{\pm i\p c} \O^\pm_R [ \z].\label{FtoR}
\ee
We can therefore restrict our attention to  $\calf = F$ when computing the refined index.
The property (\ref{FtoR}) bears an interesting relation to $\cali$-extremization \cite{Benini:2015eyy}: provided one can show that $\O^\pm_F [\z]$ is extremized at $\z = -1$, it follows that the extremal value of $\O^\pm_F [\z]$ yields (up to a phase) the $R$-weighted index $\O^\pm_R$.

\section{General $N=2B$ superconformal sigma models}\label{Secgensigma}
In this Section we consider the superconformal index (\ref{indref}) in general $su(1,1|1)$ invariant type B sigma models. After reviewing the geometric structures a the target spaces of such models possess, we formally identify the superconformal index with the mathematical index of an elliptic complex. We argue that a powerful index theorem, whose validity in this context  will be justified  in the subsequent sections, can be used to compute  it. 

\subsection{The sigma models}
We will consider general $N=2$ supersymmetric sigma models, in which supersymmetry is realized on  so-called type B multiplets, each containing 2 real bosons, 2 real fermions and no auxiliary fields\footnote{For comparison, type 2A supersymmetry is realized on multiplets containing 2 real bosons, 4 real fermions and no auxiliary fields.} . 
We will label the bosonic fields as $x^A$ and  the fermions as $\chi^A$, $A = 1 , \ldots 2 d_\CC$, where $d_\CC$ stands for the complex dimension of the target space. As shown in  \cite{Papadopoulos:2000ka},  the general Lagrangian can contain, besides the standard sigma model part which is second order in time derivatives, an additional first order {part} describing motion in a background magnetic field $F_{AB}$ (not to be confused with the {auxiliary} magnetic fields $\calf^\pm_{AB}$ which will appear in the superconformal charges)  and is of the form
\bea 
L &=& L^{(1)} +  L^{(2)}\label{LNis2}\\
L^{(1)} &=&  A_A \dot x^A  - {i \over 2} F_{AB} \chi^A \chi^B\\
L^{(2)} &=&   \half G_{AB} \dot x^A \dot x^B + {i \over 2}  G_{AB}\chi^A \hat \nabla_t \chi^B -  {1 \over 12} \pa_{[A} C_{BCD]}  \chi^A  \chi^B  \chi^C  \chi^D\label{L2gen}\\
\hat \nabla_t \chi^A &:=& \dot \chi^A + \left( \G^A_{\ BC} + \half  C^A_{\ BC} \right) \dot x^B \chi^C.
\eea

Without going into full detail (for which we refer to \cite{Mirfendereski:2022omg}) let us summarize the target space geometry of these sigma models. The presence of  $N=2$ Poincar\'e supersymmetry in the model requires the   existence of  an integrable
complex structure $J$.
The metric $G$  and field strength $F$ should be Hermitean 
with respect to $J$:
\be 
F_{AC} J^C{}_B+F_{CB} J^C{}_A=0, \qquad	G_{AC} J^C{}_B+G_{CB} J^C{}_A=0.\label{N2conds}
\ee
The  3-form $C_{ABC}$ plays the role of a fully antisymmetric torsion tensor, 
and the K\"ahler two-form $\O_{AB} = J_{AB}$ should\footnote{A slightly weaker condition would in fact suffice \cite{Mirfendereski:2022omg}, but we will only consider the restricted class of models satisfying (\ref{Omcovconst})  here.} be covariantly constant with respect to the torsionful connection $\hat \nabla$, 
\be 
\hat \nabla_A \O_{BC} =0.\label{Omcovconst}
\ee
For a given complex structure and Hermitean metric, the connection satisfying (\ref{Omcovconst}) is unique and is called the Bismut connection (see e.g \cite{Fedoruk:2014jba} for a pedagogical discussion).  The explicit expression for the Bismut torsion $C_{ABC}$ is
\begin{equation}
C_{ABC} = - 3 J^D{}_A J^E{}_B J^F{}_C \nabla_{[D} J_{EF]} .\label{CJbism} 
\end{equation} 

The existence of an R-symmetry rotating the supercharges furthermore requires the existence of  a vector field $\r^A$ 
satisfying 
\be   \call_\r G= \call_\r J = \call_\r C =0, \qquad i_\r F=0.\label{Rconds} \ee
In particular, the first two identities imply that $\r$ is a real-holomorphic Killing vector field.

As shown in \cite{Michelson:1999zf,Papadopoulos:2000ka}, superconformal invariance requires in addition the existence of a conformal Killing vector $\xi$ which leaves the complex structure invariant and satisfies
\begin{align} 
\call_\xi G_{AB} &= 2 G_{AB}, & \call_\xi J^A_{\ \  B} &=0\\
\call_\xi C_{ABC} &= 2 C_{ABC}, & i_\xi C &= 0.\label{xiCids}
\end{align}
Furthermore, the conformal Killing vector $\xi$ should be related to $\r$ as
\be 
\r ^A = - J^A_{\ \ B} \xi^B\,,\label{rhoitoxi}
\ee
and to the  special conformal generator $K$  through
\be 
2 K =  \x^2, \qquad \xi^\flat = d K.\label{Kxisq}
\ee
Introducing complex coordinates  $(z^m, \bar z^{\bar m})$ adapted to the complex structure $J$, meaning that
\be 
J_m^{\ n} = i \d^n_m,\qquad  J_{\bar m}^{\ \bar n} = - i \d^{\bar n}_{\bar m},\label{complcoords}
\ee
the properties of the Bismut connection lead to the following identities
\be
\o_{mnp}  =  C_{mnp} =0, \qquad   
\o_{m \bar n \bar p} =  \half C_{m \bar n \bar p}. \label{OmCcoeffs}
\ee
The first identity is in fact crucial for (\ref{GsFis1}) to hold and for a commuting $u(1)$ charge $J$ to exist in these models. 
One also shows that
\be 
\r^\flat = {i } (\bar \pa -  \pa)K .\label{rhoitoK}
\ee


\subsection{Differential geometry of the target space}	\label{Secdiffgeo}
Let us clarify 	what the above conditions imply for the differential geometric properties of the sigma model  target space, making contact with known structures in differential geometry. 
Due to the integrable complex structure  $J$ the  target space has the structure of a complex manifold with Hermitian metric $G$. Furthermore, it has a conformal Killing vector $\x$ which is holomorphic and closed. The vector $\r$ is a holomorphic Killing vector constructed out of $J$ and $\x$. In addition, the $U(1)$ gauge field $A$, if present, should have a field strength which is of type $(1,1)$ and satisfy $i_\r F=0$.

While  the target space is always a complex manifold, it is in general not K\"ahler. The metric $G$ is K\"ahler if and only if in addition we have 
\be 
i_\r C =0.\label{Kahlercond}
\ee
To show this, note that from 
(\ref{Omcovconst}) one obtains the   useful identity  $i_\x d\O = - i_\r C $. 
Using this and (\ref{rhoitoxi},\ref{xiCids}) we obtain an expression for the  K\"ahler form:
\be 
2 \O =  d\r^\flat - i_\r C.\label{Omid}
\ee
From this relation we see that $G$ is K\"ahler if $i_\r C$ is closed. But then $i_\r C$ has to actually vanish, since combining the above identities we have
$2 i_\r C =  i_\x d i_\r C$.

A useful observation is that the target space metrics of this kind   have the structure  of a metric cone \cite{Gibbons:1998xa}. 
Indeed, defining a radial coordinate as
\be 
r^2 = 2 K,
\ee
one can show 
that the metric takes the form
\be 
ds^2 = d r^2 + r^2 \widetilde{ds^2} ,\label{conecoords}
\ee 
with $\widetilde{ds^2}$ the metric on the base of the cone.
The conformal Killing vector is given by
\be 
\x =   r\pa_ r .
\ee
This coordinate system makes it clear that, as a direct consequence of the dilatation symmetry,  the target spaces of interest are always {\em noncompact}. Furthermore, metric cones of the form (\ref{conecoords}) over a regular base space are typically {\em singular} at the tip of the cone. The only regular examples occur when $\widetilde{ds^2}$ is the round metric on the sphere $S^{ 2d_\CC-1}$.  Therefore, in most models of physical interest one will have to deal with  the additional complication of defining the model and the superconformal index on singular target spaces. We will comment on this issue in Section \ref{Secgen2B} below.  

The restrictions on the target space  can also be viewed as specifying geometric structures on the odd-dimensional base of the cone. Generically the base of the cone is said to possess a normal almost contact structure for which  $\widetilde{ds^2}$ is an  adapted metric \cite{Boyer:2008era}. 
In the special case (\ref{Kahlercond}) where the target space is K\"ahler, 
the base is called a Sasakian manifold, and the definition (\ref{rhoitoxi}) implies that $\r$  coincides with the Reeb vector field. 
Using (\ref{rhoitoK}) we write the K\"ahler form as 
\be 
\O = i \pa \bar \pa K.
\ee
In other words we recover the familiar property \cite{Martelli:2005tp} that the radius squared $r^2 =  2 K$ plays the role 
of the K\"ahler potential.


\subsection{Quantization}
The first step in quantizing $N=2B$ sigma models is to represent the canonical commutation relations 
\be 
[ x^A, p_B] = i \d^A_B ,\qquad \{ \chi^{{A}},  \chi^{{B}} \} = G^{{A} {B}} 
\ee
in terms of self-adjoint operators on a Hilbert space $\calh$.
	One way to represent the algebra (another one, in terms of differential forms on $\calm$, will be discussed shortly) is to take $\calh$ to be a spinor bundle over   the target space equipped with the standard inner product.  
We represent the  momentum operators  and fermions as
\be p_A = - i G^{-1/4} \pa_A G^{1/4}, \qquad \chi^A = { \g^A \over \sqrt{2}} \label{momfermrepr}\ee
where $\g^A$ are the curved-index gamma matrices.
To simplify formulas below, we will also introduce the non-Hermitean derivative operator
\be 
\tilde p_A \equiv - i  \pa_A = p_A  + {i \over 2} \G_{AB}^B.
\ee
In an $su(1,1|1)$ superconformal  sigma model, the R-charge operator and the fermionic generators take the form\footnote{We mostly follow the conventions of \cite{Mirfendereski:2022omg}, with $(R, Q^{1,2}, S^{1,2},\x, \r)$ corresponding to  $(-2 R, Q^{3,4}, S^{3,4},-2 \x, -2 \o)$ in that work. } \cite{Michelson:1999zf}  
\begin{align}
R =& - \r^A \P_A  +  i \nabla_A \r_B \chi^A \chi^B,&\P_A \equiv& \tilde  p_A - A_A  - {i \over 2} \left( \o_{ABC} - \half C_{ABC} \right) \chi^B \chi^C \nonu
Q^1 =& - \chi^A J_{A}^{\ B} \P_B  + {i \over 2}  J_{[A}^{\ \ D} C_{BC]D} \chi^A \chi^B \chi^C, &   	S^1 =&     \chi^A \r_A   \label{GNC1}\\
Q^2 =&   \chi^A \P_A - {i \over 6} C_{ABC} \chi^A \chi^B \chi^C, &  S^2 =&   \chi^A  \x_{ A} 
\label{GNC2}
\end{align}
The superconformal charges $\calg_{\pm 1/2}$ are given by the complex combinations (\ref{calgdefs}), where we set  $\o=1$.  Due to the property (\ref{calgcalq}), they take the same form as
the  combination   $\calq = Q^1 + i Q^2$  albeit with  a shift of the background gauge field by an effective gauge potential $\cala^{\pm}$,
\be 
A \to A + \cala^{\pm},\label{Ashift}
\ee
where
\be 
\cala^\pm =\mp i (\pa -\bar \pa)K = \pm \r^\flat. \label{calagen}
\ee
The   field strength  corresponding to $\cala^\pm$ is
\be
\calf^\pm
= \pm 2 i \pa \bar \pa K .\label{calfK}
\ee
In other words, $K$ is a potential for $\calf^\pm$, and in the special case that the target space is K\"ahler (i.e. $i_\r C =0$), $\calf^\pm$ is proportional to the  K\"ahler form. The explicit form of the superconformal charges is simplest in adapted complex coordinates (\ref{complcoords}). One shows  from (\ref{GNC1} - \ref{GNC2}) and (\ref{OmCcoeffs}) (see \cite{Mirfendereski:2022omg} for more details) that  they  reduce to
\be 
\calg_{\pm \half} =   \chi^{\bar m} \left( i (\tilde p_{\bar m}- A_{\bar m}-\cala^\pm_{\bar m}) + \o_{ \bar m  {\bar p} {n}} \chi^{ {\bar p}}  \chi^{ {n}}+{1 \over 8} \pa_{\bar m} \ln G \right).\label{Gscomplex}
\ee

Let us now comment on the  definition of the fermion number operator and the $u(1)$ charge $J$ appearing in  the refined indices.   The fermion number operator $F$ is defined to assign charge 1 to a fermionic creation operator and charge -1 to an annihilation operator and therefore depends on how we choose to split the fermions in creation and annihilation operators. There are in general many choices which  all lead to the same Witten parity $(-1)^F$, up to an overall sign ambiguity. 
However in our sigma models  the  complex structure provides a  canonical way of splitting the fermionic operators, 
and we will choose to view the $ \chi^{\underline{\bar m}}$ as creation operators and the $ \chi^{\underline{ m}}$ as annihilation operators.  The corresponding fermion number operator is
\be 
F = \chi^{\underline{\bar m}} \chi_{\underline{\bar m}} ={i \over 2}  J_{AB} \chi^A \chi^B + {d_{\CC} \over 2}
\ee
and satisfies, as required,
\be 
[ F, \hat \chi^{\underline{\bar m}} ] = \hat \chi^{\underline{\bar m}}, \qquad  [ F, \hat \chi^{\underline{ m}} ] = - \hat \chi^{\underline{ m}} .
\ee
One checks that the Witten parity $(-1)^F$ is   represented as the chirality operator
\be 
(-1)^F = \g^{ 2 d_\CC + 1}.
\ee
 

\subsection{Reeb-like vector and extended algebra}
From (\ref{Gscomplex}) we find that both $\calg_{1/2}$ and $\calg_{-1/2}$   carry fermion number one as anticipated in (\ref{GsFis1}). 
Following the discussion below that equation, we introduce the operator
\bea 
J&=& - R + {i \over 2}  J_{AB}  \chi^A  \chi^B\label{defJgen2B}\\
&=&  \r^A \left( \tilde p_A - A_A - {i \over 2}  \o_{ABC} \chi^B \chi^C \right) - {i\over 2} \nabla_A \r_B \chi^A \chi^B ,\label{Jopexpr}
\eea
where in the last line we used (\ref{Omid}). This operator generates an additional $u(1)$ symmetry which commutes with   all $su(1,1|1)$ generators 
and can be used to define the refined indices (\ref{indref}). 
We should also remark that $J$ has a clear physical meaning as the charge corresponding to the symmetry of the sigma model  induced by the  Reeb-like Killing vector $\r$, under which  the fields transform as
\be 
\d_R x^A = \e \r^A, \qquad \d_R \chi^A = \e \pa_B \r^A \chi^B .\label{Jtransfos}
\ee
For a derivation we refer to \cite{Mirfendereski:2022omg}, see eq. (3.53).
The c-number or zero-point  term in (\ref{defJ}) was chosen in (\ref{defJgen2B}) to be $-N/2$, so that the fermionic ground states have symmetric  $J$-eigenvalues ranging from $-N/2$ to $N/2$.  

\subsection{Auxiliary supersymmetric system and resolved target spaces}\label{Secgen2B}
The superconformal indices only make use of the $N=2$ Poincar\'e subalgebra of the full symmetry which is generated by
$\calg_{\pm 1/2}, \calg_{\pm 1/2}^\dagger, \calh_\pm$  and, in the case of the refined index, of the additional central $u(1)$ generator $J$.  We recall our  observation (\ref{Ashift})  that the superconformal charges $\calg_{\pm 1/2}$ and Hamiltonians $\calh_\pm$ are obtainable  from the Poincar\'e supercharge $\calq$ resp.   Hamiltonian $H$ of the original system  (\ref{L2gen})  by a shift of the background gauge field
\be 
A_A \to  A_A + \cala^\pm_A \equiv \widetilde{A}_A^{\pm}.\label{gaugeshift}
\ee
The Lagrangians $\call_\pm$ obtained from the Legendre transform of $\calh_\pm$   describe an auxiliary system, and are related to  the original sigma model Lagrangian  (\ref{LNis2}) by the same shift (\ref{gaugeshift}) of the gauge field. This auxiliary action is  invariant under  
$N=2B$ Poincar\'e supersymmetry and under the $u(1)$ generated by $J$ (cfr. (\ref{Jtransfos})). Such sigma models are required to satisfy the following set of  conditions \cite{Mirfendereski:2022omg}, namely
\begin{align}
	G_{AC} J^C{}_B+G_{CB} J^C{}_A=&0, &
\hat \nabla_A \O_{BC} =&0, & \widetilde{F}^\pm_{AC} J^C{}_B+\widetilde{F}^\pm_{CB} J^C{}_A=&0 \label{N2Bcondsa}\\
\call_\r G = \call_\r J = \call_\r C =& 0 , & & &i_\r \widetilde{F}^\pm =& d v^\pm , \label{N2Bconds}
\end{align}
To verify that these are satisfied in the auxiliary models  with Lagrangians $\call_\pm$, only the conditions involving $\widetilde{F}^\pm$ are not automatic.  The last identity in (\ref{N2Bcondsa}) holds because
$\calf^\pm_{AC}J^C_{\ B} + \calf^\pm_{CB}J^C_{\ A}=0$ due to  (\ref{calfK}), and the last identity in  (\ref{N2Bconds}) holds with $v^\pm = - i_\r \cala^\pm$. 
%
However, as expected since $\calh_\pm$ do not commute with $R$, the auxiliary system is no longer R-symmetric   and therefore also no longer $su(1,1|1)$ invariant. Concretely  this follows because $\calf^\pm$ fails the requirement   $i_\r \calf^\pm=0$ (cfr. (\ref{Rconds})), since 
$i_\r \calf^\pm = \mp \half \x^\flat \neq 0$. 

To summarize, we can view the superconformal index as the standard Witten index of an $N=2$ supersymmetric system whose Lagrangian is obtained from (\ref{L2gen}) by the substitution (\ref{gaugeshift}).
Let us now return to the issue of defining the superconformal index  for  target spaces which are singular cones. Suppose we can find a continuous family  of  sigma models in which the  target geometry is deformed near the tip of the cone so as to resolve the singularity,   while preserving the geometric structure (\ref{N2Bcondsa},\ref{N2conds}).  
The Witten index of the resolved models is then a  natural candidate for  the superconformal index on the singular space. 
 For the subclass of models where the cone is actually K\"ahler,  resolutions of the singularity were discussed  in \cite{Martelli:2005tp}, though in the most general case this issue deserves further study and is left for future work (see \cite{p2} and the Discussion). We will proceed under the  assumption that  a sensible resolution does indeed exist.  
In what follows we will therefore study the refined Witten index on general smooth (but not necessarily compact) sigma models satisfying (\ref{N2Bcondsa},\ref{N2Bconds}) and argue that it  
is given by a simple fixed-point formula. 

 
 \subsection{Index and elliptic complex}
 
 	In supersymmetric quantum mechanical sigma models with compact target, it is typically possible to represent the supercharges as elliptic differential operators acting on some bundle over the target manifold.
 	This  allows for the identification of a Witten-type  index of the quantum mechanical model with an index (in the mathematical sense) of an   elliptic complex. In this section we will see how the superconformal index $\O_F^\pm$ can, at least formally, be interpreted as the index of an elliptic complex
 	in two equivalent ways.
 	 Similarly, the refined index $\O_F^\pm [\zeta ]$  can  formally be identified with a character-valued index of the same complex. These identifications hold at the formal level only due to the non-compactness of the target spaces of interest,  and  we will say more on this issue in the next subsection. 


 Let us first discuss the most straightforward connection to mathematical index theory, which identifies the superconformal index $\O_F^\pm$ with the index of a generalized Dirac operator. We recall that anti-chiral primary states are annihilated  by $ \calg_{1/2}, (\calg_{1/2})^\dagger$, while chiral primaries states are annihilated  by $ \calg_{-1/2}, (\calg_{-1/2})^\dagger$. An equivalent statement is that they
 are annihilated by the Laplacians $\D_\pm$ given by
\be 
\D_\pm = \left( - i \left( \calg_{\pm \half}-  \calg_{\pm \half}^\dagger \right) \right)^2.
\ee
From our previous discussion we know that the combination $ - i ( \calg_{\pm 1/2}-  \calg_{\pm 1/2}^\dagger )$ takes the form of the Poincar\'e supercharge $Q^2$ (see (\ref{GNC2})), where the background gauge field is shifted as $A \to \widetilde{A}^\pm = A + \cala^\pm $. Using (\ref{momfermrepr}) we find that this operator is represented on the spinorial  Hilbert space  as
\bea
 - i \left( \calg_{\pm \half}-  \calg_{\pm \half}^\dagger \right)&=& - {i \over \sqrt{2}} \g^A \left(\pa_A +{ 1\over 4} \left( \o_{ABC} - {1 \over 6} C_{ABC}\right) \g^B \g^C - i \widetilde{A}^\pm \right)\nonu
 & \equiv & -  {i \over \sqrt{2}} \slashed{D}^{\rm tors}_{\widetilde{A}^\pm}
\eea  
As the notation indicates, we recognize in this operator the Dirac operator  with respect to a  torsionful connection (the torsion being given by $C_{ABC}$) and twisted by a gauge field $ \widetilde{A}^\pm$. Using the fact that $(-1)^F$ is represented as the chirality operator $\g^{2 d_\CC +1}$, we can formally identify the superconformal index as an analytic Dirac index counting chirality-weighted harmonic spinors,
\bea
\O_F^\pm &=& {\rm dim \ H{arm}}^+ \left( i \slashed{D}^{\rm tors}_{\widetilde{A}^\pm}\right) - {\rm dim \ Harm}^- \left( i \slashed{D}^{\rm tors}_{\widetilde{A}^\pm}\right)\\
&\equiv& {\rm ind} \left( i \slashed{D}^{\rm tors}_{\widetilde{A}^\pm}\right),\label{Omdiracind}
\eea
where the plus (minus) superscript refers to positive (negative) chirality spinors respectively.

 Similarly it is possible to identify the refined superconformal index $\O^\pm_F [\z]$ with a character-valued index (see \cite{Eguchi:1980jx} for a review) of the same Dirac operator. Indeed, taking  the chemical potential to be a pure phase, $\z = e^{i\m}$, the index (\ref{indref}) contains an insertion of the
 $U(1)$ group element \be g_\pm = e^{\pm i\m J},\ee where $J$ is the charge associated to the Killing vector $\r$ of the target space satisfying (\ref{N2Bconds}).
 This symmetry generator $J$ acts on the   spinors as \cite{Mirfendereski:2022omg}
 \be 
 J =- i \left[ \r^A \left(  \pa_A + {1 \over 4} \o_{ABC}\g^B \g^C - i \widetilde{A}^\pm_A \right) + {1 \over 4} \nabla_A \r_B \g^A \g^B - i v^\pm\right].
 \ee
 Note that this reduces to (\ref{Jopexpr}) in the undeformed superconformal case. 
 In the combination in the square brackets we recognize the (gauge-covariantized) spinorial Lie derivative. 
Since, as one can check explicitly, the operator $J$ commutes with $ \slashed{D}^{\rm tors}_{\widetilde{A}^\pm}$, the group element $g_\pm= e^{\pm i \m J}$  has a well-defined action on the space of harmonic spinors.  
It is straightforward to see that the refined index 
$\O^\pm_F [e^{i \m}]$ becomes equal to the character-valued index
\bea 
\O_F^\pm [e^{i \m}] &=&   {\rm char_{g_\pm} \, Harm}^{+}\left(  i \slashed{D}^{\rm tors}_{\widetilde{A}^\pm} \right)-  {\rm char_{g_\pm} \, Harm}^{-}\left(  i \slashed{D}^{\rm tors}_{\widetilde{A}^\pm} \right)\\
&\equiv& {\rm ind_{g_\pm}}  \left( i \slashed{D}^{\rm tors}_{\widetilde{A}^\pm} \right). \label{GindDirac}
\eea
 Here, the notation char$_g V$ denotes  the trace of the matrix representing the action of $g$ on  the vector space $V$.

The above representation of the index involved a Dirac operator defined with respect to a torsionful connection. 
We will now work out a  simpler representation in terms  of  
a twisted Dolbeault operator.
 This is possible thanks to the special properties (\ref{OmCcoeffs}) of the  Bismut torsion tensor $C_{ABC}$, which are in fact instrumental in the representation of the Dolbeault index in terms of supersymmetric quantum mechanics on non-K\"ahler manifolds \cite{Smilga:2011ik}.
To make the connection explicit we introduce an equivalent description of the  Hilbert space   
 as the space of $(0, \bullet)$ polyforms (i.e. linear combinations of $(0,q)$ forms for various $q$)  with the standard sesquilinear inner product. 
 The fermionic operators are now represented as
 \be 
  \chi^{\bar m} = d\bar z^{\bar m} \wedge, \qquad  \chi^{  m} = g^{m \bar n} {\d \over \d ( d \bar z^{\bar n})}.\label{formrep}
 \ee
One then shows \cite{Mirfendereski:2022omg} that the superconformal charges (\ref{Gscomplex}) are represented by the following differential operators
 \bea
 \calg_{\pm \half} 
  &=& \bar \pa - i \left( \widetilde{A}^\pm_{0,1}   + {i\over 8} \bar \pa     \ln G \right)  \wedge\\
 &\equiv& \bar \pa_{V^\pm} , \qquad V^\pm =   \widetilde{A}^\pm_{0,1} +  {i\over 8} \bar \pa   \ln G   
  \eea
 where $G \equiv \det G_{AB}$ and $\bar \pa_V$ denotes the Dolbeault operator  twisted  by the gauge field $V$. Since the fermion number $F$ acts as the form degree in the representation (\ref{formrep}), it follows that the superconformal index $\O_F^\pm$  can be written as
\bea 
\O_F^\pm &=& \sum_{r=0}^{d_\CC} (-1)^r  {\rm dim \ H{arm}}^{(r)}\left( \bar \pa_{V^\pm} \right)\\
&\equiv& {\rm ind}  \left( \bar \pa_{V^\pm} \right).
\eea

In this representation  we can also relate the refined index $\O_F^\pm [e^{i \m} ]$ to a character-valued index for the differential operator $\bar \pa_{V^\pm}$. 
Using the identities (\ref{OmCcoeffs}) 
 and   (\ref{Omid}) one shows that the operator $J$ (cfr. (\ref{Jopexpr})) acts as 
\be
J = - i \left(  \call_{\r} - i\,  i_{\r} \left(\widetilde{A}^\pm - {i \over 8} (\pa - \bar \pa) \ln G \right) - i v^\pm \right)+ d_\CC. 
\ee
This gives a well-defined action on the sections of the appropriate bundle which commutes with $\bar \pa_{V^\pm}$.  The refined index can be written as a character valued index
\bea 
\O_F^\pm [e^{i \m}] &=& \sum_{r=0}^{d_\CC} (-1)^r  {\rm char_{g_\pm}\, H{arm}}^{(r)}\left( \bar \pa_{V^\pm} \right)\\
&\equiv& {\rm ind_{g_\pm}}  \left( \bar \pa_{V^\pm} \right). \label{Ginddolb}
\eea

Summarized, we have argued that a superconformal index can be viewed as an analytic index associated to an elliptic operator in two equivalent ways, namely as associated to a torsionful Dirac operator  or to a twisted Dolbeault operator.  This generalizes the well-known relation between the Dirac and the twisted Dolbault complex, see e.g. \cite{Ivanov:2010ki}, \cite{Smilga:2011ik}. 

\subsection{Superconformal index theorems}
In the previous subsection we have formally identified the superconformal index with an analytic index of a certain elliptic operator. Index theorems, which relate an analytic index with a topological index, can be a powerful tool for their computation. Such theorems are  classically derived for  complexes   associated to differential operators which possess the `Fredholm'   property that the spaces of harmonic forms are   finite dimensional. This is in particular the case for elliptic operators defined on compact manifolds. In the cases that interest us, the target spaces are non-compact and, as we will see below, the relevant  differential operators are not Fredholm.  Nevertheless, we will argue from a  physical perspective that a version of the index theorem is still applicable to compute the refined superconformal index. Mathematical work on the generalization of  the relevant index theorem to the non-compact setting appears in \cite{2000math.ph..11045B} (see also \cite{vergne2006applications} for a review).

Let us first   examine the unrefined indices and focus on  $\O^+_F$ for definiteness. It gets contributions from anti-chiral primaries, which need to be annihilated by  $\calg_{ 1/2}$ and $ \calg_{ 1/2}^\dagger$. Let us use the representation (\ref{formrep}) on $(0,\bullet)$ forms. 
Since $\bar \pa \widetilde{A}^+_{(0,1)}=0$,    we  can  locally write
\be 
- i \widetilde{A}^+_{(0,1)} = \bar \pa \l.
\ee
We then  observe that the differential equations $\calg_{ 1/2} \o= \calg_{ 1/2}^\dagger \o =0$  for anti-chiral primaries are solved by zero-form lowest Landau level wavefunctions  in  the effective magnetic background $ \widetilde{A}^+_{(0,1)} + i/8 \bar \pa \ln G$, i.e.
\be
\o_f = f(z) e^{ - \l -  {1\over 8} \ln G}\label{LLL}
\ee
where $f(z)$ is an arbitrary holomorphic function.
From (\ref{calagen}) we see that the exponential factor behaves for large $r$  as $e^{ -  K} =e^{-{ r^2 / 2}}$ and therefore provides a damping factor
rendering these states normalizeable. This heuristic argument (which we will verify  in examples in Section \ref{Secfreepart}) shows that $\O^+_F$ is expected to be divergent due to an infinite number of  anti-chiral primaries contributing with the same parity\footnote{Also in   Section \ref{Secfreepart}, we will see that the index $\O_R^\pm$ assigns different parities to the anti-chiral primaries and yields an indefinite alternating sum.}.
Let us compare this  to a naive application of the Atiyah-Singer index theorem, which   would predict the result (see e.g. \cite{Eguchi:1980jx})
\be 
\O_F^\pm = \int_\calm  e^{{\widetilde{F}^\pm \over 2 \p} + { i\over 8 \p} \pa \bar \pa  \ln G}  {\rm Td} (T_\CC M ).\label{AS}
\ee 
 Since $\widetilde{F}^\pm$ behaves at large $r$ like the K\"ahler form of the metric (\ref{conecoords}), there is a divergent contribution going like the volume of the target space.
Therefore  the divergent  topological index does seem to accurately capture the divergent analytical index in this case.  

Now let us consider the refined index $\O_F^\pm [\z ]$. On physical grounds this is expected to be better behaved since the number of lowest Landau level   wavefunctions  (\ref{LLL}) at fixed $J$-charge is typically finite, and, under  reasonable assumptions on growth of the number states at large $J$-charge, should yield a convergent power series for small $|\z |$. This leads one to speculate that the relevant  index theorem might  be applicable  in this situation. This theorem is the so-called Atiyah-Bott fixed-point formula \cite{Atiyah-Bott} and expresses the index in terms of   the $U(1)$ group action generated by $\r$ near   its fixed points. In the unresolved target space of the superconformal sigma model, the vector $\r$ has a single isolated fixed point at $r=0$, since 
\be 
\r^2 = \xi^2 = { r^2 }.\label{rhosq}
\ee
Assuming for simplicity\footnote{While this is the expected situation, this assumption is not  essential, since the result (\ref{fixedpform}) can be generalized  to the situation with several isolated fixed points or even fixed submanifolds, see e.g. \cite{Alvarez-Gaume:1986ggp}.} that the action of $\r$ on the resolved target space still has a single isolated fixed point, the  Atiyah-Bott fixed point formula predicts (see 
\cite{Eguchi:1980jx} for a detailed derivation)
\be 
\O_F^\pm [e^{i\m}] =\left. {{\rm char}_{g_\pm} E_+ - {\rm char}_{g_\pm} E_- \over \det ( 1- g_\pm (TM) )}\right|_{\r = 0}  = \prod_{n=1}^{d_\CC} {i \over 2 \sin \left( \pm {\m \o_n \over 2}\right)}.\label{fixedpform}
\ee
The `exponents' $\o_n$ in the last line  are essentially the (integer) charges of the $U(1)$ representation on the tangent space at the fixed point.  In practice they are determined as follows. Due to the fact that $\r$ is a Killing vector, it is straightforward to see that the matrix $\pa_A \r^B|_{\r =0}$ generates an orthogonal transformation (with respect to the metric at the fixed point)    and can therefore be brought into a canonical form which defines the exponents $\o_n$ in (\ref{fixedpform}):
\be 
  {\pa_A \r^B}_{| \r =0} \simeq \left(\begin{array}{ccc} 0 & \o_1 & \ldots \\-\o_1 &0 \\ \vdots& & \ddots  \end{array}\right).\label{nablarho}
 \ee
In what follows we will provide evidence for the validity of the fixed-point formula (\ref{fixedpform}) for the superconformal index, both by explicit verification in tractable examples and by giving a path-integral argument using supersymmetric localization, which does not assume compactness.  

\section{Supersymmetric localization: regular target spaces}\label{Secfreepart}
In this section we will justify the fixed point formula (\ref{fixedpform}) in the simplest examples where the superconformal sigma model has a regular target space. We will do so both by explicitly constructing the (anti-)chiral primary states contributing to the index, as well as by using supersymmetric localization of the path integral. The latter computation will pave the way for the computation in more general  (resolved) target spaces in the next section.

As already mentioned in Section \ref{Secdiffgeo}, the only superconformally invariant $N=2B$ models with regular target space metrics are in fact flat, $G_{AB} =\d_{AB}$. The complex structure is the standard 
one on $\CC^{ d_\CC}$  and the 
 torsion tensor $C_{ABC}$ vanishes.
 In adapted complex coordinates $z^m, m = 1, \ldots, d_\CC$  we have
\be 
ds^2 = d z^m d\bar z^m, \qquad  
\x = z^m \pa_m + \bar z^m \pa_{\bar m}, \qquad \r = i\left( z^m \pa_m - \bar z^m \pa_{\bar m} \right). \label{targetCd}
\ee
In principle, we should also allow for a background  gauge field $A_A$ satisfying (\ref{N2conds},\ref{Rconds}), but these conditions in fact restrict to a trivial gauge field in this case.  The gauge field should be of the form
\be 
A = A_m (\bar z) d z^m + \overline{ A_m} ( z ) d \bar z^m
\ee
The condition $i\r F =0$ can be seen to require $A_m (\bar z)$ to be linear,
\be 
A_m (\bar z)= B_{m  n} \bar z^n,
\ee
with $B$ a Hermitean matrix, $B = B^\dagger$. But such gauge fields are easily seen to be   pure gauge and lead to $F=0$.

Having established this, it is straightforward to see that the  most general regular superconformal model factorizes into  $d_\CC$ decoupled sigma models, each defined on a single complex plane, with the geometric structures  in (\ref{targetCd}) restricted to a single term in the sum.  Each of these factors allows for its own refinement associated to the rotation generator in that plane, and the total index will depend on $d_\CC$ chemical potentials and is simply given  by the product of the indices of the decoupled systems.  

\subsection{Superconformal charges}
With the above motivation we consider the free sigma model on the complex plane with  bosonic coordinate $z, \bar z$	 and associated complex fermion $\chi, \bar \chi$. The Hilbert space is $L^2 (\RR^2) \otimes \CC^2$, on which the canonical (anti-)commutation relations,  
\be 
[z, p_z] =[ \bar z , p_{\bar z}] =i, \qquad \{ \chi, \bar \chi \} = 2,
\ee
 are realized by the operators
\begin{align}
p_z &= - i \pa_z, &   p_{\bar z} &= - i \pa_{\bar z}\\
\chi &= {1\over \sqrt{2}}(\s^1 + i \s^2)   , & {\bar \chi} &=  {1\over \sqrt{2}}(\s^1 - i \s^2) 
\end{align}
It is straightforward to show that the expressions (\ref{GNC2}) for the $su(1,1|1) \oplus u(1)_J$ charges reduce to  
\bea
H &=& {2 p_z p_{\bar z}}, \qquad K =  {z \bar z\over 2}, \qquad D =-\half (z p_z + \bar z p_{\bar z} )+ {i \over 2}\label{Ncharges1}\\
L_0 &=& p_z p_{\bar z} + {z \bar z \over 4}\\
R &=& - i (z p_z - \bar z p_{\bar z} )-  \half  [\chi , \bar \chi], \qquad F= {\bar \chi \chi \over 2}\label{Rfree}\\
 J &=&  i (zp_z - \bar z p_{\bar z}) + {1\over 4} [\chi, \bar \chi ]\label{Jfree} \\
\calq &=&  { i }  \bar \chi p_{\bar z} , \qquad
\cals =    { i \over 2} \bar \chi  z   \label{Ncharges2}
\eea
The superconformal charges $\calg_{\pm 1/2}$ and operators $\calh_\pm$ then take the form 
\bea 
\calg_{\pm \half} &=& {i  }  \bar \chi \left(p_{\bar z} - \cala^\pm_{\bar z}\right)\label{calgsA} \\
\calh_\pm &=&   \left( \left(p_z -\cala^\pm_z\right)\left(p_{\bar z} - \cala^\pm_{\bar z}\right) + {\rm h. c.} \right) + i    \calf^\pm_{z \bar z} [\chi,  \bar \chi ] .\label{Hcheck}
\eea
 Here, $\cala^\pm$ is the  potential  of an  effective magnetic field  and is given by (cfr. (\ref{calagen})): 
\be
\cala^\pm = \pm {i\over 2} \left( z d\bar z - \bar z dz\right).
\ee
We see from (\ref{calgsA},\ref{Hcheck}) that the $N=2$ subalgebra generated by supercharges  $ \calg_{\pm 1/2},  (\calg_{\pm 1/2})^\dagger$  and the Hamiltonian $\calh_\pm$    
takes the form of the well-known supersymmetric Pauli system describing planar motion of a charged fermion in a 
perpendicular magnetic field (see e.g. \cite{2012supersymmetric}).
The subalgebras labelled by opposite  signs are related by flipping  {the orientation of} the magnetic field. We should also note that in the case of interest the auxiliary magnetic field is constant  and proportional to the K\"ahler form,
\be
\calf^\pm  = \pm  i  dz\wedge d\bar z .\label{freepF}
\ee
From these observations we conclude that the superconformal index {of the 2d free particle} is {the} Witten index {of} an auxiliary  Pauli system with constant magnetic field strength. Thanks to this  magnetic field the gapless continuous spectrum of the original Hamiltonian $H$, which would lead to aforementioned infrared problems when computing the supersymmetric index based on the supercharge $\calq$, is replaced by a discrete spectrum of Landau levels. We note that the idea of lifting the continuous spectrum of the original Hamiltonian $H$ by introducing a magnetic field already appears in \cite{akhoury1984anomalous}.


\subsection{Short multiplet spectrum and indices}
In this simple model we can explicitly work out the  wavefunctions for the short   multiplets  and directly compute the various superconformal indices (see also Appendix B of \cite{Benini:2015eyy}). 
We note that the R-charge operator $ R$ in (\ref{Rfree})  has integer eigenvalues, so that we can consider   indices weighted by  R-parity $(-1)^{ R}$ as well as fermion parity $(-1)^{ F}$. 
From (\ref{calgsA}) it is straightforward to  find an orthogonal basis of (anti-)chiral primary wavefunctions $\chi_{n }$ ($\tilde \chi_{n }$) for $n \in \NN$ which are annihilated by $\calg_{- 1/2},\calg_{-1/2}^\dagger$  (resp. $\calg_{ 1/2},\calg_{ 1/2}^\dagger$):
\be 
\chi_{n } = \bar z^n e^{-{ z \bar z\over 2}} \left( \begin{array}{c} 0 \\ 1\end{array} \right), \qquad  \tilde \chi_{n } =  z^n e^{- { z \bar z\over 2}} \left( \begin{array}{c} 1 \\ 0\end{array} \right).\label{wavefctnshort}
\ee
As a consistency check, one sees that these indeed {saturate the bound} (\ref{unbound}):
\be 
2   L_0 \chi_{n } =   R  \chi_{n } = (n+1) \chi_{n }, \qquad 2   L_0\tilde  \chi_{n } =-  R \tilde  \chi_{n } = (n+1)\tilde  \chi_{n }.\label{chqn}
\ee

From these results we find the spectrum of short multiplets with fixed ($(-1)^F$ or $(-1)^R$) parity in this model to be 
\begin{align} 
	N^{F\ \rm odd}_{\rm chiral} (h)&= N^{F\ \rm even}_{\rm anti-chiral} (h) = \sum_{n=0}^\infty \d_{h, {n+1\over 2}}, &  N^{F\ \rm even}_{\rm chiral} (h)&= N^{F\ \rm odd}_{\rm anti-chiral} (h) =0 \\
	N^{ R\ \rm odd}_{\rm chiral} (h)&= N^{ R\ \rm odd}_{\rm anti-chiral} (h) = \sum_{n=0}^\infty \d_{h, {n+\half}}, &  N^{R \ \rm even}_{\rm chiral} (h)&= N^{ R\ \rm even}_{\rm anti-chiral} (h) =\sum_{n=0}^\infty \d_{h, {n+1}}.
\end{align} 
Consequently, we find that the unrefined indices $\O^\pm_F$ are infinite while $\O^\pm_F$ are indefinite alternating sums:
\be 
\O_F^\pm = \pm \infty, \qquad  , \qquad \O_R^\pm = 1- 1+ 1-1 + \ldots  .\label{indexfree}
\ee

In view of these divergences, which are due to the   infinite number of  contributing short multiplets, it is  useful to consider the refined indices introduced in (\ref{indref}) which keep track of the $J$-charge of the (anti-)chiral primaries. These are  well-defined because the  number of (anti-)chiral primary states at fixed $J$ is finite (and does not grow with $J$). Indeed, from (\ref{Jfree})
 we compute the $ J$-eigenvalues of the (anti-) chiral primary states,
\be 
J \chi_{n } = - \left( n + \half \right) \chi_{n },\qquad
J \tilde \chi_{n } =  \left( n + \half \right) \tilde \chi_{n }.\label{Jchargesfree}
\ee
From these we compute the refined indices, which for $|\z|<1$  can be resummed as
\be 
\O^\pm_F [\z] =\pm {\z^{\half} \over 1-\z}.\label{indexreffree}
\ee
As a check we see that, as $\z \to 1$, these tend to   (\ref{indexfree}) for $\O_F^\pm$. Furthermore, we observe that the above result agrees with the fixed point formula (\ref{fixedpform}), where one checks from (\ref{targetCd}) that in this case the exponent $\o_1 =1$,  even though the target space is non-compact and the relevant operators fail to be Fredholm.

To obtain the refined index weighted by $\calf = R$ instead, we should continue $\z \to -\z$  and use (\ref{FtoR}) with $c= -1/2$. In particular, this gives a particular regularization   for the alternating sum in  (\ref{indexfree}):
\be 
\O_R^\pm = \mp i \O_F^\pm [-1] = \half.
\ee
The  $F$-indices $\O^\pm_F$ are regularizations of the quantity	$\tr (-1)^F$, and similarly  $\O^\pm_R$  give a meaning to $\tr (-1)^R$. We note that, while  $\O^\pm_F$  differ by a phase, the $\O_R^\pm $ agree. This reflects the fact that the fermion number $F$ cannot be defined unambiguously (it depends on an arbitrary split in fermionic creation and annihilation operators), while the R-charge $R$ is defined unambiguously. The ambiguity in the fermion number is related to a global anomaly, as explained in \cite{Hori:2014tda}.

\subsection{Path integral computation of the indices}
We now proceed to compute the superconformal indices in this simple model  using the path integral representation and standard localization techniques, and check the results against our formulas (\ref{indexfree}, \ref{indexreffree}) obtained by direct computation.
\subsubsection{Unrefined index}
Let us first consider the unrefined index $\O_F^\pm$, which we found  to be infinite as it receives contributions from an infinite number of lowest Landau level states  in a noncompact space with constant magnetic field. As we shall presently see, in the path integral computation the infinities arise from  a bosonic zero mode living on the noncompact  the target space. Anticipating this, we consider a regularized  system where we take the target space to be a torus $z \sim z+ L$, and take $L \to \infty$ in the end. Applying the  Atiyah-Singer index theorem (\ref{AS}) to this compactified system 
gives the index as   the first Chern class
\be 
\O_F^\pm =  c_1 [ \calf^\pm ] =  {1 \over 2 \p}   \int \calf^\pm .  \label{indexPauli}
\ee

The result (\ref{indexPauli}) for the Witten index of the Pauli system can also be derived from the functional integral representation of the index  using supersymmetric localization \cite{Alvarez-Gaume:1983zxc,Friedan:1983xr}. For what follows it is instructive to repeat this derivation in this simple setting, essentially following \cite{Alvarez-Gaume:1986ggp}. 
Let us therefore consider  a generic Pauli system of the form (\ref{calgsA},\ref{Hcheck}), where the background gauge field $\cala$ is arbitrary.
Performing the Legendre transform we obtain the canonical Lagrangian  
\bea
\call  &=& \half \dot z \dot {\bar z} +\cala_z \dot z +\cala _{\bar z}  \dot{\bar z}  + {i\over 2} \bar \chi \dot \chi+ {i }\calf_{z \bar z} \bar\chi  \chi
\eea
This Lagrangian is invariant, up to total derivatives, under  the $N=2$ supersymmetry  variations\footnote{We follow the conventions of \cite{Mirfendereski:2020rrk} for variations and Poisson brackets in systems with fermions.} generated by a supercharge $\calg$ of the form (\ref{calgsA}), defined as $\d_\pm  F \equiv \{ F, \calg_{\pm 1/2} \}$ for any phase-space function $F$, namely
\begin{align}
	\d_\pm  z =&0, &
	\d_\pm  \bar z =& { i  }\bar \chi\nonu
	\d_\pm  \chi =&   \dot z,&
	\d_\pm  \bar \chi =& 0\label{susy}
\end{align}
Note that these transformations are identical to the ones generated by the original complex supercharge $\calq$. This is because, in the Hamiltonian formalism, only the combination $p_{\bar z} - \cala^\pm_{\bar z} = \dot z$, which is independent of $\cala^\pm$,  appears  on the right hand side of the variation.
One checks that the action is indeed invariant and leads to a Noether charge 
which agrees with the phase-space expression (\ref{calgsA}).

We can now write a path-integral expression for the Witten index $\O_F = \tr (-1)^F e^{- \b \calh}$: 
\be 
\O_F  = \int [D\bar zD z D \bar\chi D  \chi ]_{\rm PBC} e^{- S_E}.\label{OmPI}
\ee
Here,  $S_E$ is the Euclidean action obtained  {from} $\call$ upon continuing $t \to -i \t$,
\be 
S_E = \int_0^\b d\t \left( \half \dot z \dot{\bar z}-i (\cala_z \dot z +\cala_{\bar z} \dot{\bar z}) + \half \bar \chi \dot \chi - {i }\calf_{z \bar z} \bar \chi  \chi\right)\label{euclactionfreea}
\ee
The path integral measure in (\ref{OmPI}) integrates over fields periodic in $\t$ with period $\b$, which for the fermions amounts to  inserting  $(-1)^F$. With these boundary conditions, the action is invariant under the Euclidean continuation of the supersymmetry transformations  (\ref{susy}).

We know from the Hilbert space interpretation or, alternatively, from expressing the action as a supersymmetric variation, that the index (\ref{OmPI}) is actually independent of $\b$.   To make the $\b$-dependence explicit,  we work with a rescaled time coordinate $\tilde \t = \t/\b$, so that the action takes the form
\be 
S_E = \int_0^1 d\tilde \t \left( {\b^{-1}\over 2} \dot z \dot{\bar z}-i (\cala_z \dot z +\cala_{\bar z} \dot{\bar z}) +\half  \bar \chi \dot \chi - {i \b}\calf_{z \bar z} \bar \chi  \chi\right),\label{euclactionfree}
\ee
where the dot now stands for the derivative with respect to $\tilde \t$.
In the $\b \to 0$ limit, the path integral localizes on constant  bosonic configurations. In accordance with general expectations for supersymmetric localization,   these are also precisely the configurations for which the supersymmetric variation of the fermions in (\ref{susy}) vanishes.   Expanding the bosonic integral in fluctuations around constant configurations, 
\be
z (\tilde \t)= z_0 + \sqrt{\b} \d z (\tilde \t) \label{bosfluctfree}
\ee 
the path integral simplifies further as $\b \to 0$. The second term in (\ref{euclactionfree}) is subleading, and the last term is subleading  for the nonconstant modes of the fermion fields, though for the constant fermionic mode it is the leading contribution and should be kept. To leading order in $\b$ we therefore find\footnote{Our path integral measure is normalized such that
	$\int [DxD\chi] e^{-\half x^T \calo_B x - \half \chi^\T \calo_F \chi} = \left({\det \calo_F \over \det \calo_B  }\right)^\half$, for a symmetric operator $\calo_B$ and antisymmetric operator $\calo_F$.}
\be 
\O_F = \det\;\hspace{-3pt}'( - \b^{-1} \pa_{\tilde \t})^{-1} {\b \over 2\p} \int \calf,\label{indexundrefflat}
\ee
Here, the prime on the determinant means that we omit the constant mode, and the $\b$-dependence in the determinant comes from the Jacobian of the transformation (\ref{bosfluctfree}). 
It remains to evaluate the functional determinant, which we zeta-regularize to find
\be 
\det\;_{\hspace{-3pt}\rm PBC}' (-\b^{-1} \pa_{\tilde \t}) =    \prod_{n\in \ZZ \backslash \{0\}} \left( - { 2\p i  n \over \b}\right)=  \prod_{n\geq1} \left( { 2 \p n \over \b}\right)^2 := \b .\label{zetareg}
\ee
Substituting in  (\ref{indexundrefflat}) we reproduce (\ref{indexPauli}). We note that the path integral expression has a sign ambiguity from the ordering in the fermionic measure, which corresponds to the ambiguity in the definition of the fermion number operator in the operator formalism (and which was chosen to match the conventions there). To obtain the superconformal indices $\O_F^\pm$ for the  particle on the complex plane, the field strength should  be taken to be  constant (see (\ref{freepF})), and the  constant mode $z_0$ is a bosonic zero mode.  As we let the size $L$ of the system tend to infinity, the zero-mode integration  leads to the infinite result (\ref{indexfree}) for $\O_F^\pm$.

\subsubsection{Refined index}\label{SecPIfree}

Now we turn   to the path integral computation of the refined index. Thanks to the relation (\ref{FtoR}) we can restrict   our attention 
to the index graded by the fermion number $\O_F^\pm [\z]$.   We will compute it  in the regime where $\z$ is a phase and is related to a chemical potential $\m$ as
\be 
\z = e^{ i \m}.
\ee
To read off the spectrum of (anti-) chiral multiplets (\ref{indrefmults}) we should  analytically  continue the result  and determine the  coefficients in a power series  expansion around $\z =0$.

To proceed, we seek a path-integral representation for  
\be 
\O_F^\pm [e^{ i \m}] = \tr (-1)^F e^{\pm i \m J} e^{ - \b  \calh_\pm} .
\ee
There are in fact two natural ways to translate this into path-integral language, both of which we will discuss here. The first way comes from observing that the insertion of the $U(1)$ group element $g_\pm =e^{ \pm i \m J}$ can be implemented  in the path integral by imposing $g_\pm$-twisted boundary conditions on the fields. Since both $z$ and $\chi$ carry unit charge under $J$ (see (\ref{Jfree})), these boundary conditions read
\be 
z(\t + \b ) = e^{ \pm i \m}z(\t) , \qquad \chi (\t + \b ) =   e^{\pm i \m} \chi(\t).\label{twistedbc}
\ee
The path integral formula for the index is 
\be 
\O^\pm_F [ e^{ i \m}] = \int [DzD\bar z D \chi D \bar \chi ]_{\pm \m } e^{- S^\pm_E},\label{OmPI2}
\ee
where the integral is taken over fields obeying the twisted boundary conditions (\ref{twistedbc}), and the Euclidean action is given by (\ref{euclactionfree}) with $\cala$ replaced by $\cala^\pm$. We again use $\b$-independence of the index to compute it for $\b \to 0$. The calculation proceeds as in the previous section,  the main   difference being that   the twisted boundary conditions  (\ref{twistedbc}) do not allow for constant modes of the fields. Following the same steps  as in the previous subsection leads as in (\ref{indexundrefflat})  to the functional determinant
\be 
\O^\pm_F [ e^{ i \m}]  = \det\,_{\hspace{-3pt}\pm \m} ( -   \pa_{ \t})^{-1}.\label{dettwistedbcfree}
\ee
Here, the subscript $\pm \m$ on the determinants indicates that the operator acts on functions satisfying the boundary conditions (\ref{twistedbc}).  Under these boundary conditions the eigenvalues of the operator $- \pa_\t$ are 
\be 
- i {2 \p n \pm \m \over \b},
\ee 
and we find 
\be
\O^\pm_F [ e^{ i \m}]  =\mp { i \m \over \b}\prod_{n\geq1} \left( { 2\p n \over \b}\right)^2 \prod_{m\geq1} \left( 1 - \left( { \m \over 2 \p m}\right)^2\right).
\ee
The  first, $\m$-independent infinite product diverges and can be zeta regularized as in (\ref{zetareg}) to give a factor of $\b$. The second infinite product converges and is equal to $2 \sin { \m \over 2} /\m$. Putting this together we find
\be 
\O_\pm [ e^{ i \m}] = \pm \left( e^{ - { i \m \over 2}}- e^{  { i \m \over 2}} \right)^{-1},\label{Omtwistedbcfree}
\ee
in perfect agreement with our direct computation (\ref{indexreffree}) and with the fixed point formula (\ref{fixedpform}). As already remarked at the end of the previous subsection, both calculations involve an overall sign choice, related to the ambiguity in the fermion number $F$, which cannot be fixed unambiguously.  

We now discuss a second way  of
computing the index which   is similar to the  proof of the Atiyah-Bott fixed point formula (\ref{fixedpform}) using supersymmetric localization  
	\cite{Alvarez-Gaume:1983zxc}  (see \cite{Alvarez-Gaume:1986ggp} for a review). 
	This method makes it clear  that the path integral localizes on the fixed point of the Killing vector $\r$, namely the origin $z=0$, and depends only on the $\r$ action   near this fixed point. In this way of computing the index we treat the chemical potential term as a deformation of the Hamiltonian
\be 
\calh_\pm \to   \calh_\pm^\m = \calh_\pm \mp {i \over \b } \m J.
\ee
We are therefore led to a path integral with periodic, untwisted boundary conditions but with a modified action:
\be 
\O_F^\pm [ e^{i\m}]= \int [DzD\bar z D \chi D \bar \chi ]_{\rm PBC} e^{- S^\pm_{E,\m}}.\label{OmPIref}
\ee
From Legendre transforming $ \calh_\pm^\m $ and continuing to Euclidean time we find this deformed action to be 
\bea 
S_{E,\m}^\pm &=& \int_0^1 d\tilde \t \left( {\b^{-1}\over 2}\left|  \dot z \pm i \m z \right|^2 + \half \bar \chi \dot \chi \pm {i \over 2} \m \bar \chi \chi\right.\nonu
&&\left. -i (\cala^\pm_z \dot z +\cala^\pm_{\bar z} \dot{\bar z})   - {i \b}\calf^\pm_{z \bar z} \bar \chi  \chi\right),\label{defactionfree}
\eea
where we have rescaled $ \t = \b \tilde \t$ to make the $\b$-dependence explicit.  As a consistency check, one can verify   that this modified action is still invariant under the Lagrangian version of the  supersymmetry variations generated by $\calg_{\pm 1/2}$, which read 
\begin{align}
	\d_\pm  z =&0, &
	\d_\pm  \bar z =& { i  }\bar \chi\nonu
	\d_\pm  \chi =&   {i\over \b} \left( \dot z \pm i {\m } z\right),&
	\d_\pm  \bar \chi =& 0.\label{susyref}
\end{align}
For $\b\to 0$, the bosonic integral
localizes on configurations satisfying
\be 
\dot z = \mp i \m z,
\ee
which are again those for which the fermionic variations vanish.
The only such configuration compatible with the periodic boundary conditions is in fact $z(\tilde \t) =0$, i.e. the fixed point of the rotational Killing vector $\r$. Expanding in bosonic fluctuations
\be 
z(\tilde \t) = 0 + \sqrt{\b} \d z(\tilde \t), \label{fluctreffree}
\ee
we see that only the terms in the first line of (\ref{defactionfree}) contribute in the $\b \to 0$ limit. Performing the  Gaussian integrals, taking into account the Jacobian from (\ref{fluctreffree}),  we obtain 
\bea 
\O^\pm_F [ e^{i\m}]  &=& \det\;_{\hspace{-3pt}\rm PBC} \left( -  \left( \pa_{ \t}\pm i{ \m \over \b}\right)\right)^{-1}\\
&=&  \det\;_{\hspace{-3pt}\pm \m} ( -   \pa_{ \t})^{-1}.
\eea
Hence we find  agreement with the previous computation method (\ref{dettwistedbcfree}), leading again to the result (\ref{Omtwistedbcfree}). 

 \section{Supersymmetric localization: general $N=2B$ models}\label{SecPI}
 We now turn to the path-integral computation of the refined Witten index for general $N=2B$ sigma models   possessing a $U(1)_J$ symmetry commuting with the superalgebra. The target space satisfies the conditions (\ref{N2Bcondsa},\ref{N2Bconds}). As argued in Section \ref{Secgen2B}, such an index is  expected to capture the superconformal index on general singular cones upon resolving the singularity at the tip. 
 The computation  requires a slight generalization of Alvarez-Gaum\' e's proof of the Atiyah-Bott fixed point formula \cite{Alvarez-Gaume:1983zxc}, and provides a justification for  it's validity for the non-compact target spaces under consideration.
 \subsection{Unrefined index}
 Let us first briefly comment on the unrefined index $\O_F^\pm$, which is formally given by the Atiyah-Singer index theorem (\ref{AS}). This result can also   be justified using supersymmetric localization 
   following the same philosophy as in the  simple example above, and involves localizing the path integral as $\b \to 0$. This computation involves a generalization of the original works \cite{Alvarez-Gaume:1983zxc,Friedan:1983xr} where the target space was assumed to be K\"ahler, and the terms in (\ref{LNis2})  involving the  torsion tensor $C_{ABC}$ were absent.
This generalization was performed in detail in \cite{Smilga:2011ik}. As discussed in that work, since the fermionic constant modes are strongly coupled, the four fermion term contributes and could in principle combine with higher loop contributions from the bosons at the leading order in $\b$. An additional supersymmetric deformation argument is needed to show that the torsion terms do not contribute to the index     
and that the end result is given by (\ref{AS}).

 	  \subsection{Refined index}
 We now turn to the path-integral representation of the refined index.	Following our discussion  in Section \ref{SecPIfree} we again  compute the refined indices $\O^\pm_F [\z ]$ in the case that $\z$ is a pure phase $\z = e^{i \m}$:
 	\be 
 	\O_F^\pm [e^{i \m}] = \tr (-1)^F e^{ - \b  \calh_\pm^\m } ,
 	\ee
 	where
 	 \be 
   \calh_\pm^\m = \calh_\pm \mp {i \over \b } \m J.
 	\ee
 	 We will follow the second method of computation explained  in Section \ref{SecPIfree} and treat the $\m$-dependence as a deformation of the Hamiltonian. This method again emphasizes the localization of the path integral on the fixed point of the Killing vector $\r$ and the fact that it is fully determined by the corresponding $U(1)_J$ action in the vicinity of the fixed point. 
 We  are therefore led to a path integral with periodic  boundary conditions on the fields but with a $\m$-dependent action:
 	\be 
 	\O_F^\pm [ e^{i\m}]= \int [DzD\bar z D \chi D \bar \chi ]_{\rm PBC} e^{- S^\pm_{E,\m}}.\label{OmPIref2}
 	\ee
 		From Legendre transforming $ \calh_\pm^\m $ and continuing to Euclidean time we find the deformed action to be
 	 \bea
 	 S_{E,\m}^\pm &=& \int_0^1 d\tilde \t \left[ {\b^{-1}\over 2}G_{AB} \left( \dot x^A \pm \m \r^A\right) \left( \dot x^B \pm \m \r^B\right) - i \tilde A^\pm_A  \dot x^A  \right. \nonu
 	&&+  \half G_{AB} \chi^A \hat \nabla_\t \chi^B  \mp {\m \over 2}\left(\nabla_A \r_B + \half \r^C C_{CAB} \right) \chi^A \chi^B\nonu
 	&&\left. + {i\b  \over 2} \tilde F^\pm_{AB}  \chi^A \chi^B+  {\b \over 12} \pa_{[A} C_{BCD]}  \chi^A  \chi^B  \chi^C \chi^D\right].\label{SEmugen}
 \eea
 From the first term we see that, in the $\b \to 0$ limit, the path integral localizes on configurations with
 \be 
 \dot x^A (\tilde \t)  = \mp \m \r^A (x(\tilde \t)).
 \ee
 The only such configurations compatible with the periodic boundary conditions are constant scalars taking values at the fixed points of the Killing vector $\r$. As motivated below (\ref{rhosq}), we assume\footnote{A derivation for the case of non-isolated fixed points will appear in \cite{raeywip}.} that
the Killing vector $\r$    has a single, isolated, fixed point at $x=x_0$.
 We expand the bosons in  fluctuations   around the fixed point
 \be 
 x^A = x_0^A + \sqrt{\b} \d x^A.\label{bosflucts}
 \ee
We find that only a few quadratic terms survive in the $\b \to 0$ limit. Using the fact that $\r$ is Killing and that, at the fixed point, partial derivatives of $\r^A$ can be replaced by covariant ones we can write the contributing terms as
 \bea
  S_{E,\m}^\pm &=& \int_0^{1} d\tilde \t \left[ \half  \d x^A \left.\left(-  G_{AB}    \pa_{\tilde \t}^2 \pm 2 \m \nabla_A \r_B \pa_{\tilde \t} -  \m^2  \nabla_A \r_C  \nabla^C \r_B \right)\right|_{x_0} \d x^B \right.  \\
  && + \left. \half \chi^A \left.\left(  G_{AB}    \pa_{\tilde \t}  \mp \m \nabla_A \r_B  \right)\right|_{x_0}  \chi^B +\calo (\sqrt{\b} )  \right]
  \eea
  At this point we observe that the terms in the sigma model action (\ref{SEmugen}) involving the gauge field $\tilde A^\pm_A$ and the torsion tensor $C_{ABC}$ do not contribute to  the index. The computation therefore proceeds completely as   in \cite{Alvarez-Gaume:1983zxc,Friedan:1983xr}, which considered  K\"ahler sigma models in which those terms were absent to begin with.
  Let us nevertheless  spell out the rest of the computation for completeness.
  
  Performing the Gaussian path integral (keeping the Jacobian from the change of variables in (\ref{bosflucts})), we find a partial cancellation between bosonic and fermionic determinants and end up with
  \be
  \O^\pm_F [ e^{i \m}] = {\rm det}_{\rm PBC}\left( - \d^A_B \pa_\t \pm {2\m \over \b} \nabla_B\r^A (0) \right)^{-\half} 
  \ee
  Diagonalizing the matrix $\nabla_B \r^A$ using (\ref{nablarho}) the result can be written as a product of determinants of the type computed in Section \ref{Secfreepart}:
  \bea 
   \O^\pm_F [ e^{i \m}]  &=& \left(\prod_{n=1}^{d_\CC} {\rm det}_{\rm PBC} \left( \pa_\t^2 -  \left(\mp  { i \m \o_n \over \b}\right)^2\right)\right)^{-\half}\\
   &=&\prod_{n=1}^{d_\CC} {i \over 2 \sin \left( \pm {\m \o_n \over 2}\right)}
  \eea
  Note that, in accordance with our remarks  on the ambiguity in choosing  the fermion number operator, we have made a convenient choice for the branch of the square root. This result constitutes a justification of the index formula (\ref{fixedpform}) from supersymmetric localization.

\section{Outlook}
In this work we have initiated the study of superconformal indices for type $N=2B$ sigma models, whose    target spaces are noncompact complex manifolds with a conformal Killing vector. We have in particular constructed a refined index which  on physical grounds is expected to be  finite and well-behaved. A powerful tool for its computation is the Atiyah-Bott fixed point formula (\ref{fixedpform}), which we have justified in this non-compact setting by use of a supersymmetric localization argument. 

One point which we didn't address in detail is the proper definition and computation of the superconformal index on  target spaces  which are singular cones. A plausible construction would be to resolve the singularity at the tip of the cone    while preserving the geometric structures needed to define the refined index. The details of such a resolution and consistency checks on this procedure deserve further study.
For type $4A$ superconformal models (which are K\"ahler), such a resolution  was proposed in \cite{Dorey:2019kaf} for target spaces which can be realized as complex symplectic varieties. 
For our $N=2B$ models, whose target spaces possess less structure,   it would be  good to have a more differential geometric approach to resolving the singularity.  We plan to address this issue in a future publication \cite{p2}.


One of the motivations for this work was to pave the way for the computation of  superconformal indices  for quiver quantum mechanics in the Coulomb branch \cite{Denef:2002ru} and in an AdS$_2$ scaling limit. These can be formulated as $N=4B$ sigma models in which a $U(1)$ symmetry is gauged \cite{Mirfendereski:2022omg}, and possess $D(2,1;0)$ superconformal invariance \cite{Anninos:2013nra,Mirfendereski:2020rrk}. Study of these models is hoped to shed light on  the stringy origin  of AdS$_2$/CFT$_1$ duality.

\section*{Acknowledgements}
It is a pleasure to thank Heng-Yu Chen, Nick Dorey,  Tom\'a\v{s} Proch\'azka, Paolo Rossi and  Andy Zhao for useful discussions. The work of CS was supported by the European Union's Horizon Europe programme under grant agreement No. 101109743, project Quivers. He would like to thank DAMTP and Center for Theoretical Physics, National Taiwan University for their hospitality where part of this work was done. The research of J.R. was supported by European Structural and Investment Funds and the Czech Ministry of Education, Youth and Sports (Project FORTE CZ.02.01.01/00/22\_008/0004632).
DVdB was supported by the Bilim Akademisi through a BAGEP award. He would like to take this occasion to thank all the members of the theoretical physics community that he had the chance to interact with over the years, in particular all his coauthors and collaborators. 
 	 \begin{appendix}

	\section{No superconformal index for $N=1$}\label{AppnoindNis1}
	In this Appendix we {argue} {that} no well-defined superconformal index exists in $N=1$ superconformal models. More precisely, we will show that any candidate superconformal index  is in fact equivalent to the naive supersymmetric index and therefore suffer from the same problem related to the continuity of the spectrum of $H$.
	
	 The  $N=1$ superconformal algebra is  $osp(1|2)$ with  {the following} (anti-)commutation relations in addition to (\ref{sl2r1}) 
	\bea
\,	\{Q,Q\}&=&2H\qquad \{S,S\}=2K\qquad \{S,Q\}=-2D\\
\,	[D,Q]&=&-\frac{i}{2}Q\qquad [D,S]=\frac{i}{2}S\qquad [H,S]=-i Q\qquad [K,Q]=i S\label{slsusy}
	\eea
	To define a  Witten-type index, we start by picking a fermionic generator $\calg$ which can a priori be an arbitrary complex combination of $ Q$ and $S$.  However, a short calculation shows that, in order for a bosonic combination $\calh = a H + b D + c H$ to exist which commutes with it,  $\calg$ must be a real combination of $ Q$ and $S$ and $\calh$ must be proportional to $\calg^2$. Hence, without loss of generality we can take 
	\be 
	\calh = \cosh \a\, Q + \sinh \a\, S,\qquad  \calh = \calg^2, \qquad \a \in \RR.
	\ee
	The main observation is that $\calg$ and $\calh$  are unitarily equivalent to $Q$ and $H$ respectively:
	\be 
	\calg = U Q U^{-1}, \qquad \calh = U H U^{-1},
	\ee 
	where
	\be 
	U = e^{i \a (H - K)}.
	\ee
Therefore our auxiliary Hamiltonian $\calh$ suffers from the same problem of having a continuous spectrum extending down to zero  as the original  Hamiltonian $H$ and cannot be used to obtain a wel-defined superconformal index.

\end{appendix}                                                                                                                                                                       
\bibliographystyle{ytphys}
\bibliography{references}
\end{document}